\documentclass{aa}
\usepackage{psfig}
\newcommand{\Teff}{\mbox{$T_{\rm eff}$}}

\begin{document}

\title {Blue irregular variable stars in the Small Magellanic Cloud from EROS2 : Herbig Ae/Be or classical Be stars ?\thanks{This work is based on observations at the European Southern Observatory,
La Silla, Chile.} }
 \author { J-P.~Beaulieu\inst{1},
W.J.~de Wit\inst{1,2,3},
H.J.G.L.M.~Lamers\inst{2,3}, 
J-B.~Marquette\inst{1},
C.~Coutures\inst{4}, P.~Leisy\inst{5}, S.~Totor\inst{1}, 
N.~Palanque-Delabrouille\inst{4},
C.~Afonso\inst{4},
C.~Alard\inst{1},
J.N.~Albert\inst{6},
J.~Andersen\inst{8},
R.~Ansari\inst{6},
\'E.~Aubourg\inst{4},
P.~Bareyre\inst{4},
F.~Bauer\inst{4},
G.~Blanc\inst{4},
X.~Charlot\inst{4},
F.~Couchot\inst{6},
F.~Derue\inst{6},
R.~Ferlet\inst{1},
P.~Fouqu\'e\inst{5,10}
J.F.~Glicenstein\inst{4},
B.~Goldman\inst{4},
D.~Graff\inst{9},
M.~Gros\inst{4},
J.~Haissinski\inst{6},
J.C.~Hamilton\inst{11},
D.~Hardin\inst{4}
J.~de Kat\inst{4},
T.~Lasserre\inst{4},
\'E.~Lesquoy\inst{1,4},
C.~Loup\inst{1},
C.~Magneville \inst{4},
B.~Mansoux\inst{4},
\'E.~Maurice\inst{7},
A.~Milsztajn \inst{4},
M.~Moniez\inst{6},
O.~Perdereau\inst{6},
L.~Pr\'evot\inst{7},
N.~Regnault\inst{6},
J.~Rich\inst{4},
M.~Spiro\inst{4},                 
A.~Vidal-Madjar\inst{1},
L.~Vigroux\inst{4},
\and S.~Zylberajch\inst{4}
}

\institute {    
        Institut d'Astrophysique de Paris, 98bis Boulevard Arago, 75014 Paris, France 
        	\and
        Astronomical Institute, University of Utrecht,
        Princetonplein 5, NL-3584 CC, Utrecht, The Netherlands 
        	\and 
        SRON Laboratory for Space Research, Sorbonnelaan 2, NL-3584 CA, Utrecht, The Netherlands 
        	\and
        CEA, DSM, DAPNIA, Centre d'Etudes de Saclay, 91191 Gif sur Yvette, Cedex, France
                 \and
        ESO La Silla, casilla 19001, Santiago 19, Chile.
        	\and
        Laboratoire de l'Acc\'{e}l\'{e}rateur Lin\'{e}aire,
        IN2P3 CNRS, Universit\'e de Paris-Sud, 91405 Orsay Cedex, France
        	\and
        Observatoire de Marseille,
        2 pl. Le Verrier, 13248 Marseille Cedex 04, France
		\and     
Astronomical Observatory, Copenhagen University, Juliane Maries Vej 30,
2100 Copenhagen, Denmark
		\and      
       Departments of Astronomy and Physics, Ohio State University, Columbus,
OH 43210, U.S.A.
                \and
       DESPA, Observatoire de Paris section Meudon, 1 place Jules Janssen, Meudon cedex F-92195, France 
                \and 
       Coll\`ege de France, 11 Place Marcelin-Bertherlot, 75231 Paris cedex 05 
}
\authorrunning{Beaulieu et al.}
\titlerunning{ Small Magellanic  Cloud blue irregular variables}
\offprints {    J.P. Beaulieu, beaulieu@iap.fr}
\date{Received date; accepted date}

\abstract{ Using data from the EROS2 microlensing survey, we report
the discovery of two blue objects with irregular photometric behaviour
of $\Delta V~\sim~0.1$-$0.4~$mag on time scales of 20 to 200 days.  They show a bluer
when fainter behaviour.  Subsequent spectra taken with the ESO 3.6m telescope
show spectral type B4eIII and B2eIV-V with strong $H \alpha$
emission.   These objects resemble the Herbig AeBe but also classical Be stars.
At this stage, it is not possible to distinguish unambiguously between pre-main sequence
and classical Be nature. If we favour the  pre-main sequence interpretation,
they are more luminous than the luminosity upper limit for Galactic HAeBe
stars. The same was found for the HAeBe candidates in the LMC.  This might
be due to a shorter accretion time scale ($\tau = M_*/\dot{M}$), or
the smaller dust content during the pre-main sequence evolution of SMC
and LMC stars.  Further studies on a larger scale of the environment, 
and IR properties of the stars are needed.
 }
\maketitle

%

\section{Introduction}

We present the discovery and the study of two Small Magellanic Cloud
blue irregular variable stars.  We noticed that during EROS-2 SMC data
base mining for microlensing by Palanque Delabrouille et al. (1998)
10 irregularly blue variable objects were detected (see their Sect.\,3 to describe
the method).  Among them, two have the same variability
characteristics as the Galactic Herbig Ae/Be stars (Herbig 1960, Bibo
\& Th\'{e} 1991, Waters \& Waelkens 1998 and references therein), and
were chosen for deeper analysis, similar to our previous studies of HAeBe
candidates in the LMC (Beaulieu et al. 1996, and Lamers et al. 1999, de Wit et
al. 2001a, BL, LBD, DBL hereafter).  We discuss their photometric and
spectroscopic properties, and we present several arguments which might
lead to the interpretation of these stars as pre-main sequence (PMS)
HAeBe stars. If confirmed, these stars would be the first SMC pre-main
sequence stars.  Following the nomenclature introduced in BL, we will
call the stars {\bf E}ROS {\bf S}MC {\bf H}AeBe {\bf
C}andidates or ESHC stars.

A large scale systematic search over the full EROS2 SMC database as
done in the LMC by de Wit et al. (2001a) will be presented in the near
future (de Wit et al. 2001b).

\section { The observations}
\subsection{Searching for blue irregular objects in the SMC} 

The field has been observed as part of the ongoing EROS-2 microlensing
survey.  The set-up consists of a 1m F/5 Ritchey-Chr\'etien telescope
with two $2 \times 4$ CCD detectors in different focal planes covering
a field of $0.7 \times 1.4 $ degree.  The light is split by a dichroic
allowing simultaneous observing in two non standard bandpasses ($V_E=$
420-720 nm and $R_E=$620-920 nm).

During the search for microlensing in the EROS2 SMC data base (5.3
million stars coming from 10 square degrees of the densest part of the
SMC) of one year observations (Palanque-Delabrouille et al., 1998), 10
blue irregular variable objects were discovered which are located in
the HRD to the right of the main sequence. Two of these displayed
irregular photometric variability with different time scales,
superimposed on long term variability. The light curves resemble those
of the HAeBe stars.  

\begin{figure*}
\vskip 9.0truecm
 \caption{The upper left panel is the finding chart (3.25 arcmin
 $\times$ 1.65 arcmin) in the $H \alpha_{off}$ band of the two targets ESHC1 and ESHC2, labeled by the left and the right
 cross respectively.  North is up, East is left.  The two other are $(H
 \alpha - H \alpha _{off})/ \sqrt{H \alpha}$ images.  These figures show
 only the $H_\alpha$ point sources and the $H_\alpha$ emitting regions
 (stripes are related to data reduction problems like bad columns or
 saturated stars).  The lower left panel is the area shown on the
 finding chart, the right panel gives the full field observed.  One can
 notice the presence of an H\,II region, and clusters of $H \alpha$
 emitters.}
\end{figure*}

The stars, called ESHC1 and ESHC2, have the following coordinates :\\
ESHC1: RA= 00:53:03.0 and  DEC= -73:17:59, J2000 \\
ESHC2: RA= 00:52:32.6 and  DEC= -73:17:08, J2000. \\
The findings charts in $V$ are given in Fig 1.

The full $V_E$ and $V_E-R_E$ light curves of theses two stars are
shown in the two upper panels of Fig 2. The two lower panels show the
low amplitude very irregular photometric behaviour in $V_E$ and $R_E$
for both stars. The uncertainty in the photometry is typically
$\sigma V_E \approx \sigma R_E\approx 0.02$ mag.

The light curves of both stars show a long time scale behaviour
(several hundred days $\Delta V_E=0.2-0.4$ mag) with superimposed
short time scale irregular variability (several days, $\Delta V_E=0.1$
mag).  {\it They both have a bluer when fainter behaviour for the long time
scale variation and the short time scale variation.}  The very
irregular photometric behaviour on short time scale is characteristic
of Herbig AeBe stars, but may also be compatible with classical Be stars.  
We will come back to this in Sect. 5.

\begin{figure*}
 \centerline{\psfig{figure=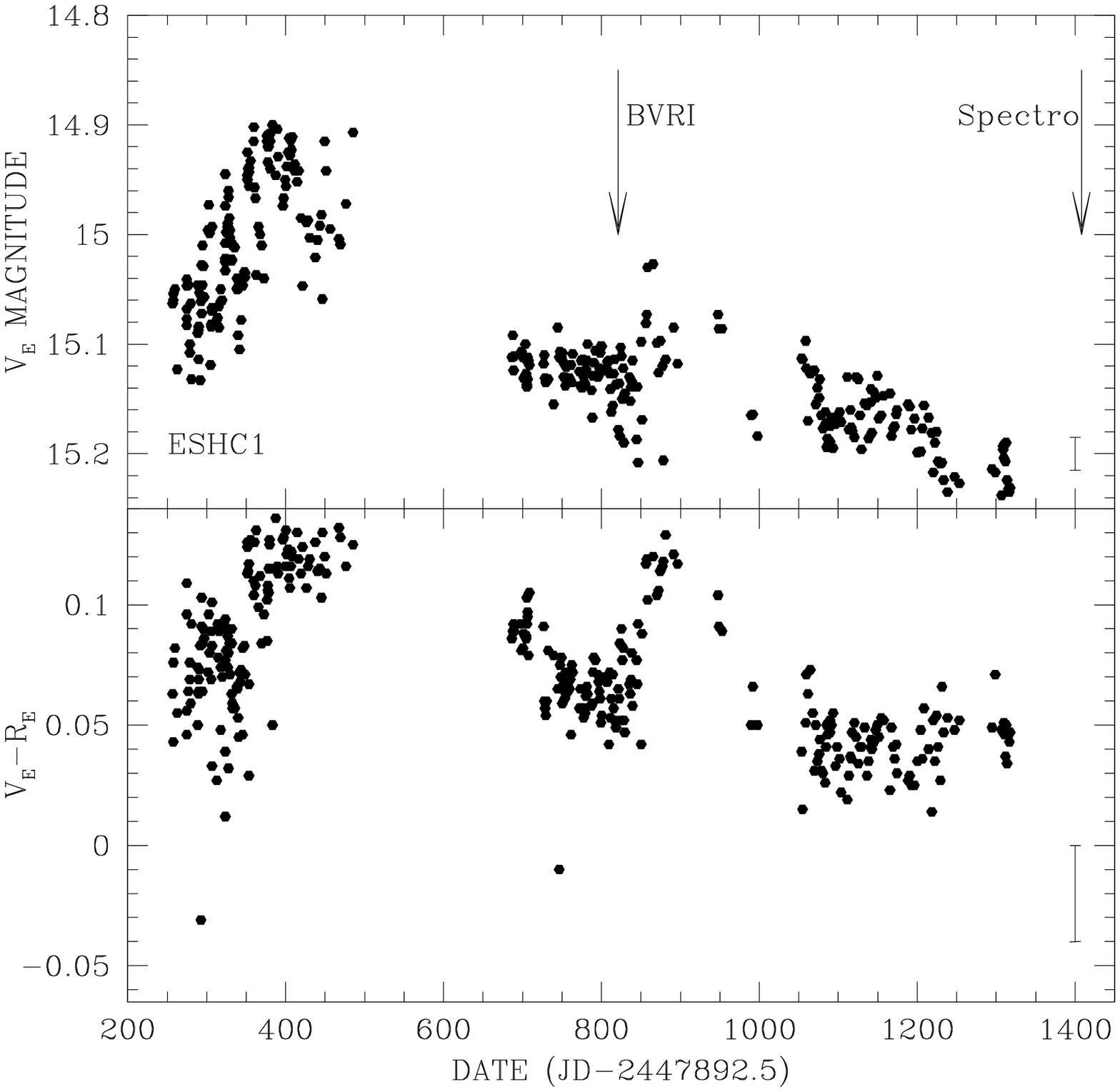,height=6cm,width=9cm}
 \psfig{figure=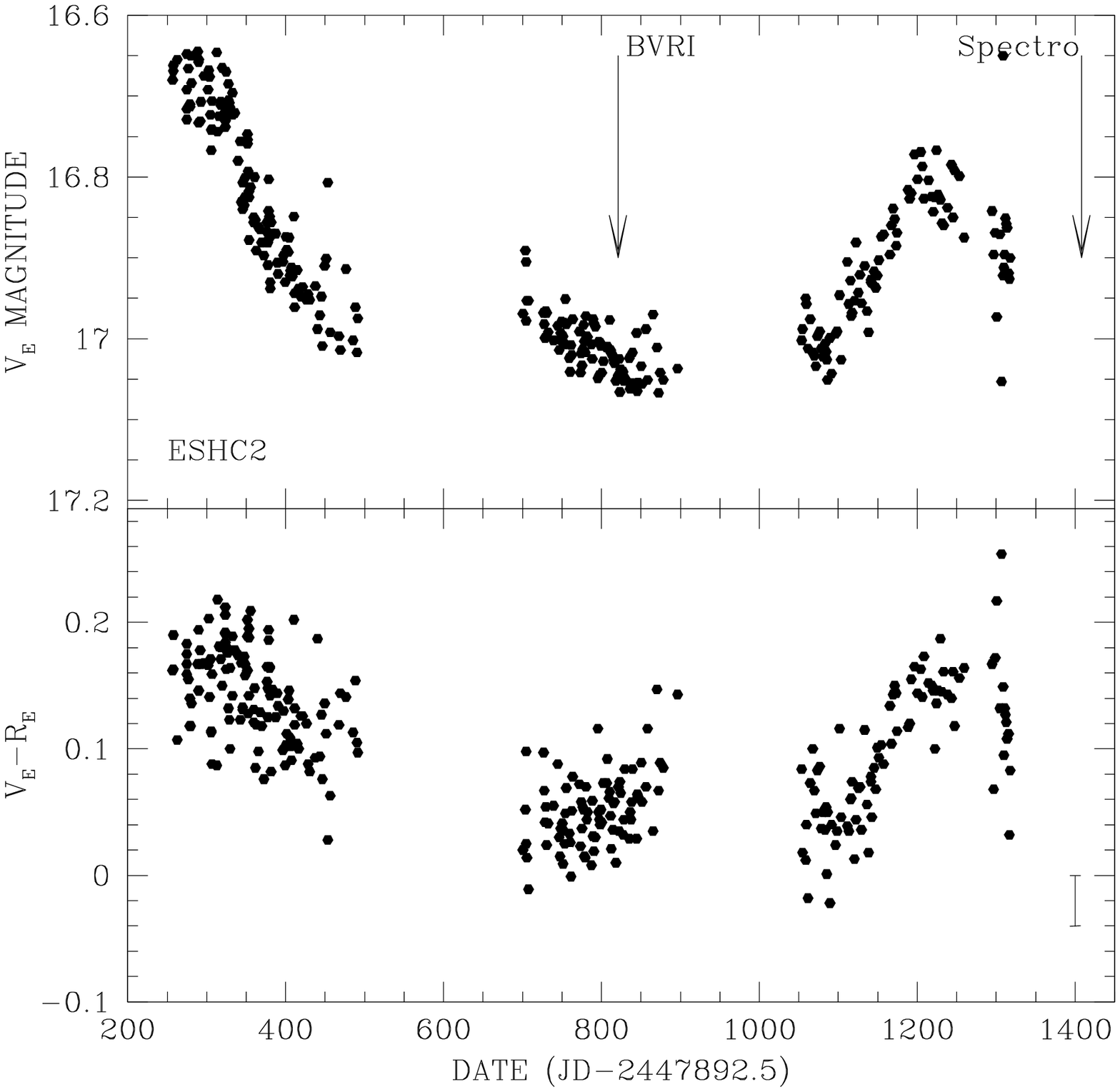,height=6cm,width=9cm}}
 \centerline{\psfig{figure=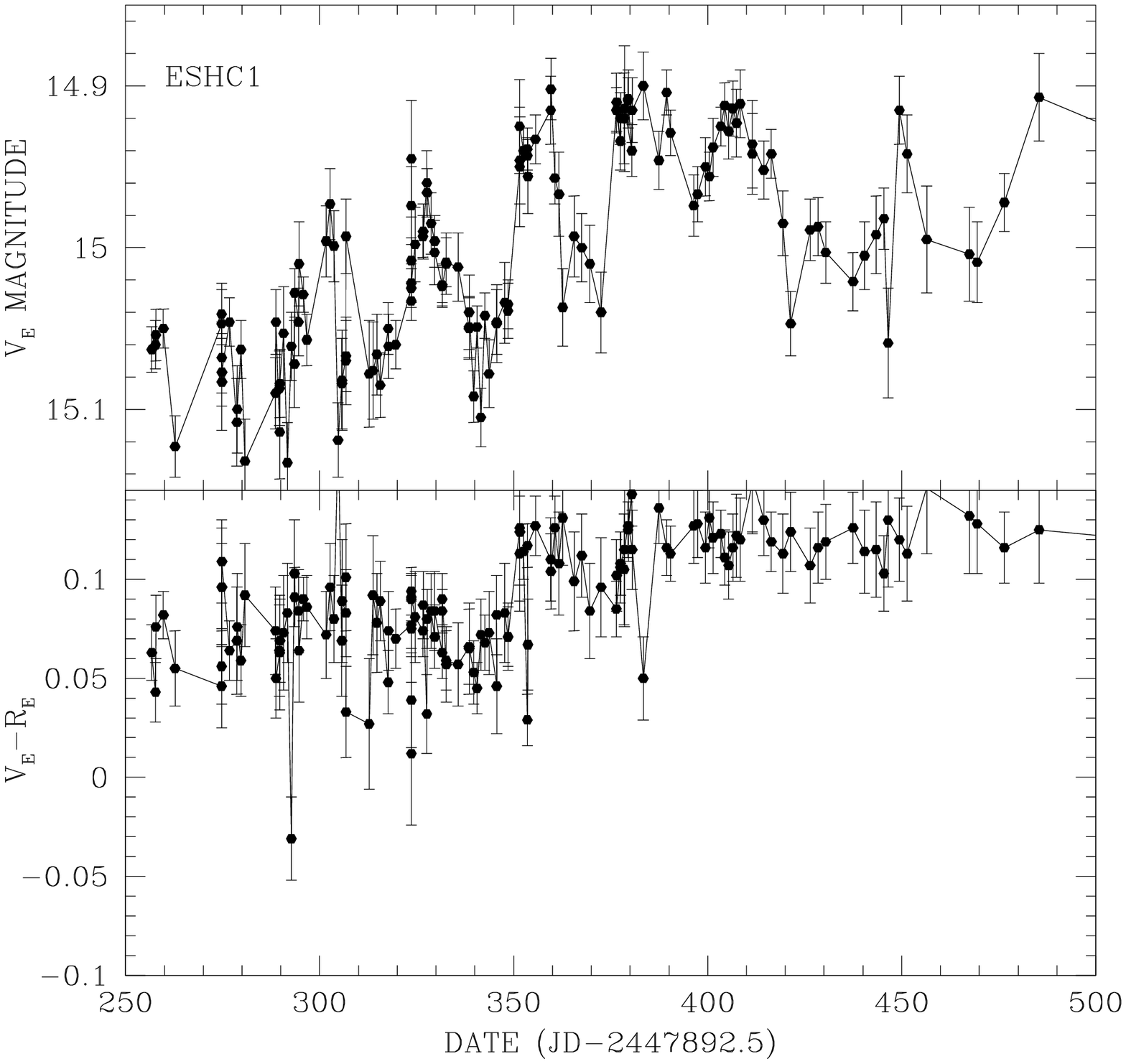,height=6cm,width=9cm}
 \psfig{figure=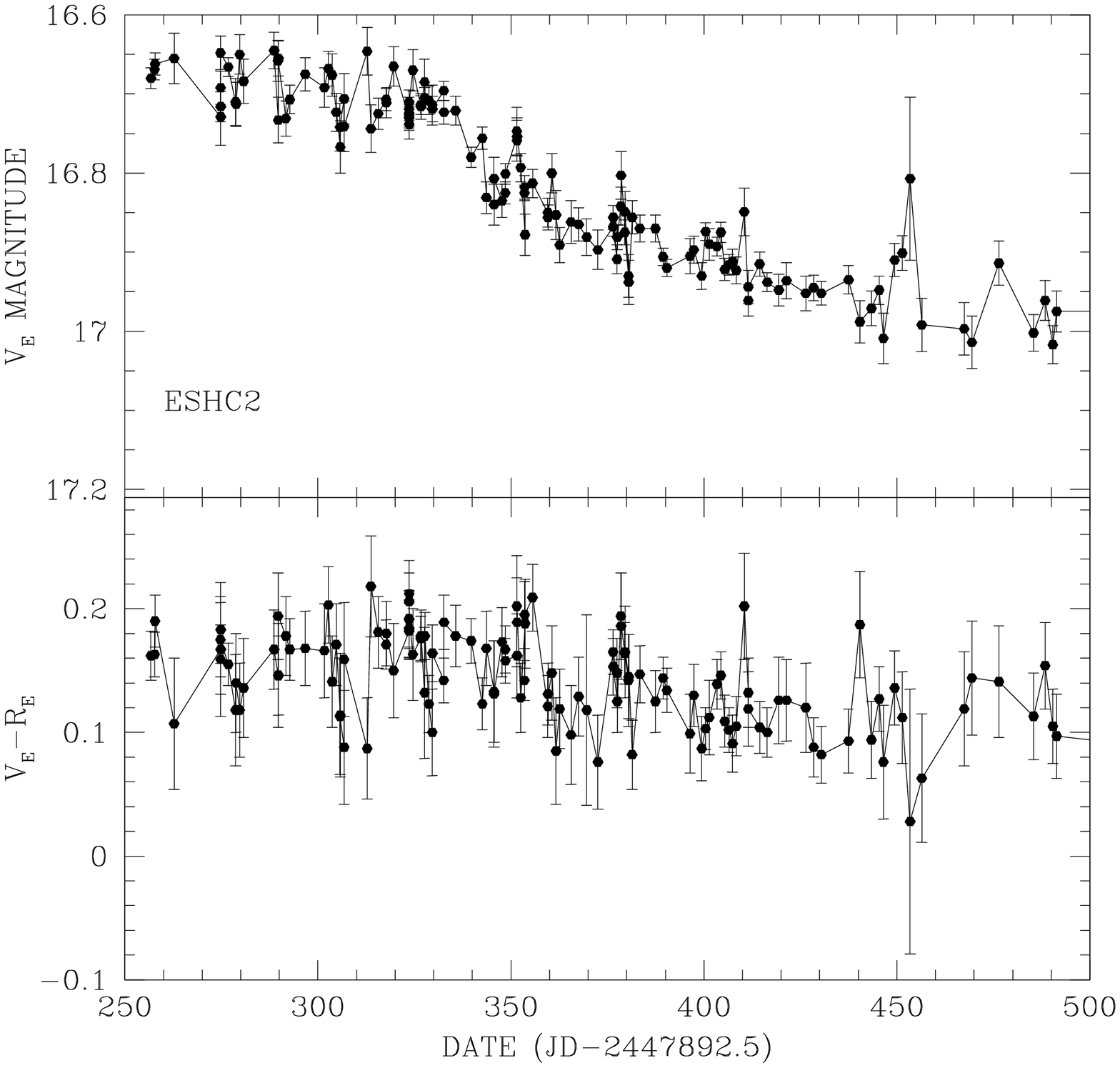 ,height=6cm,width=9cm}}
 \caption{Light curves of the two targets. The two upper panels shows
 the full $V_E$ and $V_E-R_E$ light curves for ESHC1 and ESHC2.  The
 two lower panels show zooms on a part of the light curves. Notice the
 low amplitude irregular photometric variability.  The time is given as
 JD-2447892.5. A typical error bar is given in the lower right of the two upper
 panels.}
\end{figure*}

\subsection {The photometry }

We noted that ESHC1 is catalogued by Meysonnier \& Azzopardi (1993) as
an emission line star in their objective prism survey, and
 identified as Lin 232 in the Simbad database.
Despite the fact that ESHC2 was in the field observed by Meysonnier \&
Azzopardi, it was not detected as an $H \alpha$ emitting object.

On January 7, 1998 (JD 2450821.39) we have obtained BVRI$H \alpha$H$ {
\alpha _{off}}$ photometry at ESO La Silla with DFOSC on the Danish 1.5m
for the SMC field, with exposure times of 60 sec, 30 sec, 30 sec, 30
sec, 600 sec and 600 sec respectively.  The filters are the
respectively numbered ESO 450, 451, 452, 425, 693, 697.  The CCD is a
LORAL 2048x2048 CCD with a pixel scale 0.39 arcsec.  The data have
been reduced using QUYLLURWASI, the pipeline developed around DoPhot
for the PLANET collaboration.  The absolute calibration has been done
using Landolt standards taken during the observations, leading to
absolute calibration to 2 \% for stars brighter than V=17~mag, and of
the order of 5 \% for stars of V=18 mag.

Finally, using WCSTOOLS and the MACS catalogue (Tucholke et al.,
1996), we performed astrometry of the reference frames and the
subtracted frames.  Comparison with other catalogues indicates that we
have an accuracy better than 0.5 arcsec.

We also searched for the program stars in the publicly available OGLE
BVI catalogues for the SMC of 2.2 million stars. These magnitudes are
mean magnitudes over the observing period covered by OGLE
(http://www.astrouw.edu.pl/ $\sim$ ftp/ogle/). The photometry is
summarized in Table 1.

\begin{table*}
 \begin{center}
 \begin{tabular}{llrrrrlrr}
 \hline
 Name & Vd & (B-V)d & (V-R)d & (V-I)d &$H \alpha-H \alpha _{off}$ & Vog & (B-V)og & (V-I)og  \\
 \hline
 ESHC1 & $15.09 \pm 0.02$ & $ 0.03 \pm 0.02$ & $0.06 \pm 0.02$ & $0.11 \pm 0.02$ & $-0.42\pm 0.03$ & $15.07 \pm 0.01$ & $-0.00 \pm 0.03$ & $0.13 \pm 0.04$\\
 ESHC2 & $17.03 \pm 0.02$ & $-0.03 \pm 0.02$ & $0.05 \pm 0.02$ & $0.10 \pm 0.02$ & $-0.58\pm 0.03$ & $17.01 \pm 0.02$ & $-0.05 \pm 0.03$ & $0.07 \pm 0.04$\\
 \hline
 \end{tabular}
 \end{center}
 \caption{Photometry obtained for ESHC1 and ESHC2 with DFOSC on the Danish telescope at 
 ESO La Silla on January 7, 1998 (indicated by a ``d'') and from the published catalogue 
 of average SMC BVI photometry by OGLE (indicated by a ``og'').}
\end{table*}
Moreover, narrow band imaging in $H \alpha,~H \alpha _{off}$ was obtained
with EFOSC2 mounted on the ESO 3.6m at La Silla on August 18, 1999
(Julian day 2451408.79) for the field containing the targets with
exposure times of 600 sec.  Flatfielding, bias subtraction, cosmic
subtraction were performed with MIDAS 98NOV in a standard
way. Although the seeing was poor (1.3 arcsec), these images are
useful to search for $H \alpha $ emitters in the field.  We built
subtracted images ($H \alpha-H \alpha _{off}$) using ISIS, the new
image subtraction package with non constant kernel (Alard 2000). The
procedure is the following : we choose the $H \alpha$ image as being the
template. We re-map the $H \alpha _{off} $ image by bicubic spline
interpolation from the $H \alpha$ image. Then we adjust the parameters
of ISIS to our images, and do several tests with different degrees to
fit the variation of the convolution kernel over the image. We also
performed several tests to fit the background variations. The best
results are obtained with a degree 1 variation in both the kernel and
the background.

We present images of the ($H \alpha-H \alpha _{off})/ \sqrt{H \alpha}$
flux in the lower left panel and the right panel of Fig. 1.  It gives
an idea of the noise statistics compared to the photon noise : the
gray scale has been adjusted so that black means 4 sigma above photon
noise level.  It is clear that both ESHC objects are strong $H \alpha$
emitters. They lie on a faint HII region illuminated by nearby hot
stars. Notice that at 5 arcsec (1.5 pc on the sky) to the North West
from ESHC2, there is another $H \alpha$ emitter.  Moreover,
there are several $H \alpha$ emitters clustered in the upper left
corner of the field.  This indicates that we are most likely in an area 
where star formation was active recently.


\subsection { The spectroscopic observations }

Spectroscopic observations were obtained with EFOSC2 mounted on the
ESO 3.6m at La Silla on August 18, 1999.  The gratings nr. 7 and nr. 9
were used to cover the full spectral range from 3200 to 7000 \AA. We
used a 1.2 arcsec slit, and we binned the data to a resolution of
6\AA.  Multiple exposures of the two stars were made with the grating
nr. 7 (1200 sec exposure), and grating nr. 9 (900 sec exposure) for each
star.  The spectra were reduced in a standard way using MIDAS 98NOV
release and the LONG context.  A signal to noise ratio of 200 was
obtained for ESHC1 and 50 for ESHC2 in the blue part of the spectrum.
A somewhat lower signal to noise ratio was obtained in the red part of
the spectrum: 100 for ESHC1 and 30 for ESHC2.

\section{ Discussion of the spectra}

The full spectra are presented in the Fig. 3 with ESHC1 in the upper
panel and ESHC2 in the lower panel. The region between $3800 < \lambda
< 4500~$\AA\ is enlarged in Fig. 4. The H$\beta$ and H$\alpha$ profiles
are shown in Figs. 5 and 6.  We immediately notice that these two stars
have stellar spectra with small Balmer jumps, clear Balmer lines in
absorption and strong H$\alpha$ emission. This suggests spectral type
Be.

In order to improve the estimated spectral type, we decided to adopt
the same approach as in LBD.  We first looked for the presence or
absence of classification lines, Balmer lines, HeI, HeII, CIII, OII,
MgII, SiII, SiIV. When it was possible, we measured the position and
equivalent width of the lines.  It is clear that the metal lines are
weak, as expected for metal poor stars in the SMC, and marginally
detected if detected at all.  Therefore, in the following we will
concentrate on the Balmer lines and the Helium lines. The results are
reported in Table 2.  In the next two subsections we will discuss the
spectral type of the two program stars.

\begin{table*}
 \begin{center}
 \begin{tabular}{|l|r|l|r||l|r|l|r||l|}
 \hline
 \multicolumn{4}{|c|}{Star ESHC1} & \multicolumn{4}{|c|}{Star ESHC2} & \multicolumn{1}{c|}{} \\
 \hline
 position&  FWHM &\multicolumn{1}{c|}{EW}&Flux  & position&FWHM &\multicolumn{1}{c|}{EW} & Flux &Identification\\
  \multicolumn{1}{|c|}{(\AA)}    &\multicolumn{1}{c|}{(\AA)}  &\multicolumn{1}{c|}{(\AA)}   &      & \multicolumn{1}{c|}{(\AA)}    
 &\multicolumn{1}{c|}{(\AA)} &\multicolumn{1}{c|}{(\AA)} & & \\
 \hline
3822.8   &6.0   &-0.54  &-17   & & & & &HeI 3819\\
3892.0  &10.4   &-3.2   &-106  & 3889.8 & 18.0  &-2.7    &-16  &HeI 3888\\
4012.2  & 6.    &-0.4   &-12   & & & & & HeI 4009\\
4029.4  & 7.    &-0.8   &-23   & 4028.4 &  11   &-0.8 &  -4.5    &HeI 4026\\
4146.5  & 7.6   &-0.6   &-17.5 & 4147.7 &  7  &-0.7   &-4.   & HeI 4146\\
4391.5  & 9.3   &-0.74  &-20.2 & 4392.0  & 7.6  &-0.8-1.0&   &HeI 4387\\
4475.1  & 8.8   &-0.83  &-22   & 4473.8  &6     &-0.6&  -3 &HeI 4471\\
4717.3  & 7.5   &-0.28  &-6.8   & & & & &HeI 4713\\
4925.7  & 6.8   &-0.62  &-13   & 4925.0 & 7.4  & -0.7 & -3.4  &HeI 4921\\
5019.6  & 6.2   &-0.23  &-4.65  & & & & & HeI 5015\\
6684.3  &10.9   &-0.6   &-5.6   & & & & & HeI 6678\\
\hline
5894.5  &12.6   &-1.0   &-13   & 5890.6 & 14   & -1.0 & -3.2   &  Na \\
\hline
3737.3   &6.0   &-1.1   &-33   & & & & & $H_{13}$ 3734\\
3753.0   &7.5   &-1.7   &-51   & & & & & $H_{12}$ 3750\\
3773.4   &7.9   &-2.1   &-65   &3773.1 &13 &-2 &-11 & $H_{11}$ 3770\\
3800.9   &9.12  &-2.58  &-81   & 3800 & 14 & -2.3 & -13&$H_{10}$ 3798 \\
3838.4   &8.8   &-2.5   &-81   & & & & & HeI 3838 + $H_{9}$ 3835\\
3936.8   & asym      &       &      & 3934.2    &  9.7  &-0.9    &-5.4   & Ca H\\
3972.7  & 10.4  &-3.4   &-110   & 3972.1        & 13.9  &-4.0    &-24    &H$\epsilon$ / H$\epsilon$ + K\\
4105.1 & 9.7    &-2.33  &-70   & 4105.4 & 14.1  &-2.7    &-16    &H$\delta$\\
4344.2  &12.7   &-2.54  &-69    &  \multicolumn{3}{|c|}{filled}&  &H$\gamma$\\          
4865.5  & \multicolumn{3}{|c|}{partly filled} &4864.5  & 9 & 2 & 9 & H$\beta$\\  
6567.8  & 7.5  & 6.2   & 58   & 6567.4 & 10.1 & 18.5 & 40.4   & H$\alpha$ \\
\hline
 \end{tabular}
 \end{center}
 \caption{Lines identified in ESHC1 and ESHC2. We give the measured wavelength of the lines (columns 1, 5), the 
 Full Width at Half Maximum (given in columns 2, 6), the equivalent width (column 3, 7) and the measured Flux (in columns 4, 8).}
\end{table*}

\begin{figure}
 \centerline{\psfig{figure=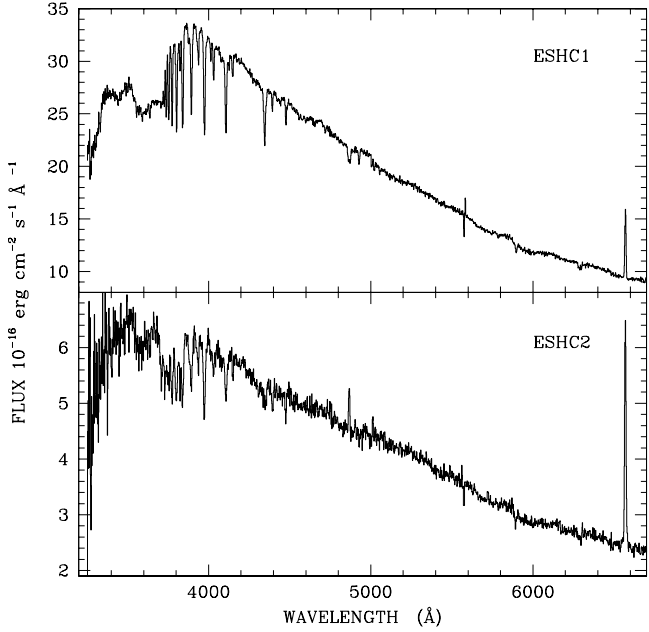,height=9cm,width=9cm}}
 \caption{Full spectrum of the two targets taken with EFOSC2 on the ESO 3.6m. The spectra have a resolution of 6 \AA\ .}
\end{figure}

\begin{figure}
 \centerline{\psfig{figure=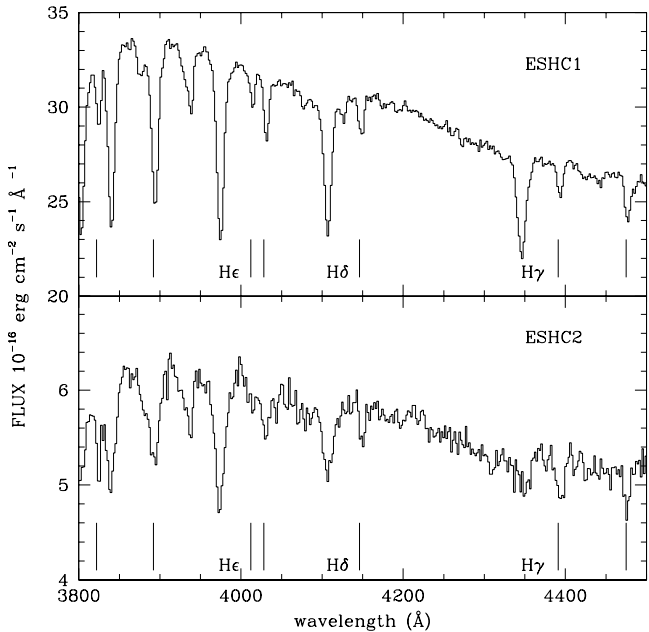,height=9cm,width=9cm}}
 \caption{The blue part of the spectrum of ESHC1 and ESHC2. Balmer lines H$\epsilon$, H$\delta$, H$\gamma$ are labeled, and
 the solid lines vertical indicate HeI lines at 3822, 3892, 4012, 4026, 4146, 4387 and 4471 \AA.}
\end{figure}

\begin{figure}
 \centerline{\psfig{figure=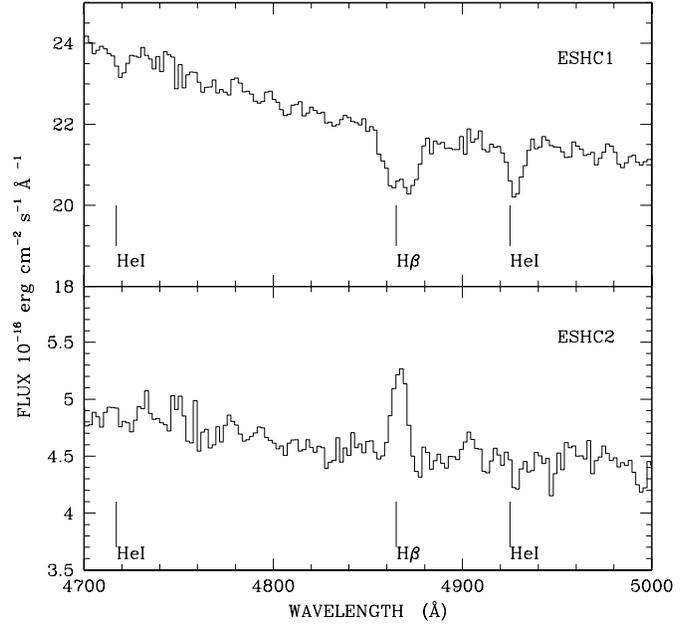,height=9cm,width=9cm}}
 \caption{The $H \beta$ line and HeI lines at 4713 \AA\ and 4925 \AA\ of ESHC1 and ESHC2.}
\end{figure}

\begin{figure}
 \centerline{\psfig{figure=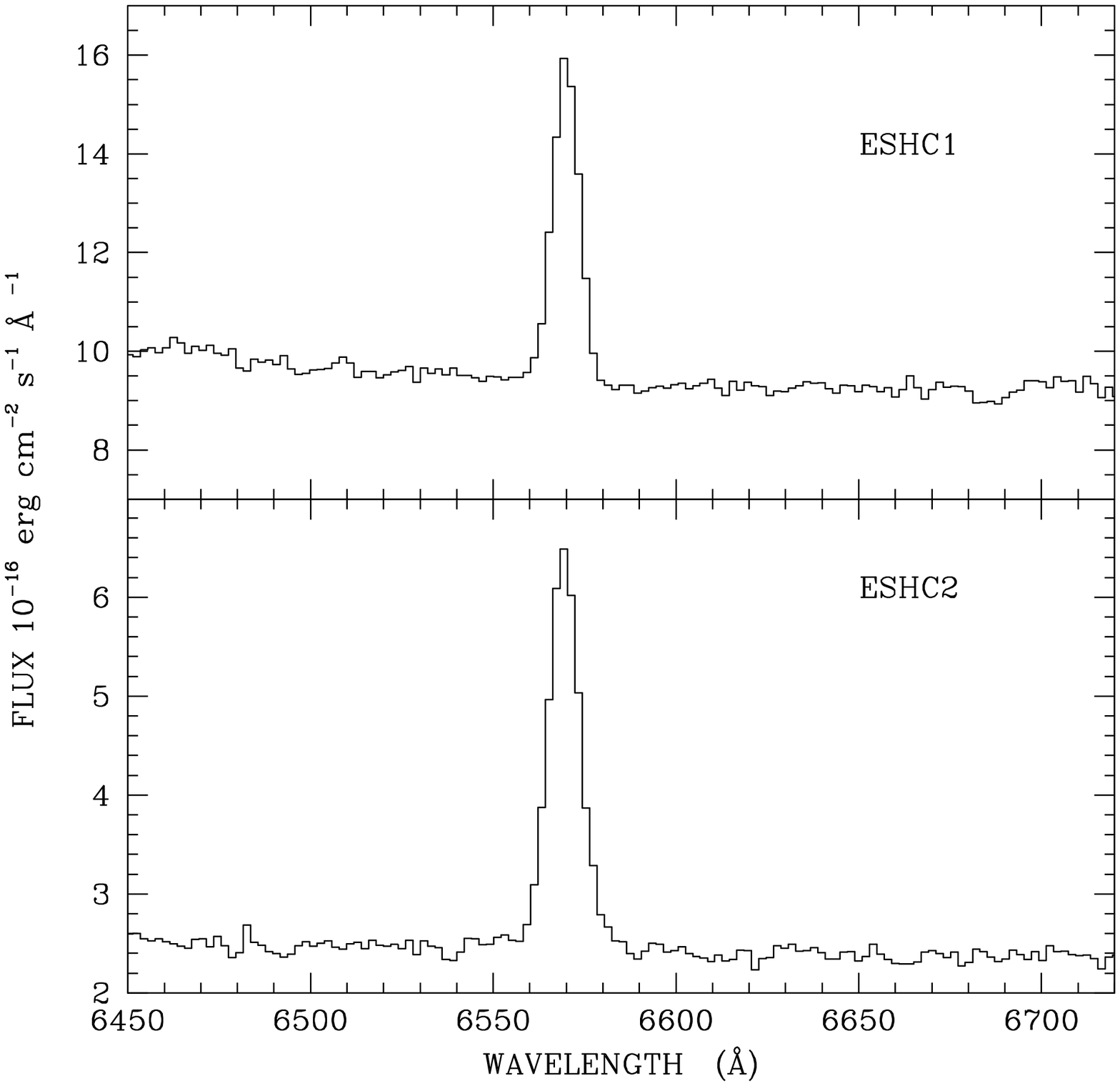,height=9cm,width=9cm}}
 \caption{The $H \alpha$ line of ESHC1 and ESHC2.}
\end{figure}

\subsection{The spectrum of ESHC1}

ESHC1 shows H$\alpha$ in emission, H$\beta$ is partly filled with
emission, all other Balmer lines from H$\gamma$ to H12 and Helium I
(3819, 3888, 4026, 4387, 4471, 4713, 4921, 5015, 6678 \AA) are
present, and clearly seen in absorption (Fig. 4).  This star has a
clear Balmer jump. The HeII line at 4686 \AA\ is absent.  There may be
a weak contribution of the CIII/OII blend at 4650, but if present, its
equivalent width would be less than 0.1 \AA.  There may be a trace of
OII at 4662 \AA. There may also be a trace of SiIV at 4116, but the
SiIV counterpart at 4089 \AA\ is in the wing of H$\epsilon$. Therefore
the presence of these lines is inconclusive.  The MgII doublet at 4481
\AA\ is not detected.

We compare the data with the tables from Didelon (1982) of the
equivalent width as a function of spectral type for several lines.
Most of the metal lines do not give strong constraints in this
particular case.  However, the presence and strength of the H and HeI
lines suggest that this star is a Be star with a luminosity class III,
IV or V.  The spectral type and luminosity class is best constrained
by the HeI 4387 line, which indicates a luminosity class III, and a
spectral type B4-6.  The HeI 4471 line also implies that the star has
a type between B4 and B6, but HeI 4026 line implies a slightly earlier
spectral range of B1 to B4.  Based on these lines, we classify this
star as a B4e III.

\subsection{The spectrum of ESHC2}

ESHC2 shows H$\alpha$ and H$\beta$ in emission.  All other Balmer
lines, from H$\gamma$ to H13, and HeI (3888, 4026, 4143, 4392, 4471
\AA) are present and clearly in absorption.  This star has a weaker
Balmer Jump than ESHC1.  The HeII line at 4686 \AA\ and the MgII
doublet at 4481 \AA\ are absent.  No features are seen at 4650 \AA\
from the CIII/OII blend.  Just like for ESHC1, we used the tables of
Didelon (1982) on the HeI lines at 4471, 4026, 4387 and 4120 \AA\ to
determine the spectral type and luminosity. Based on the comparison of
the observed line strength with Didelon's tables we classify this star
as a B2e IV-V.

\subsection{Spectral type from the Q method}

We can also estimate the spectral type from the photometry, using the
relation between the reddening-free $Q$-parameter and the spectral
type.  For this purpose we adopt the transmission curves of the $U, B$
and $V$ filters and convolve these with the observed spectra to derive
the $U$, $B$ and $V$ magnitude in the Johnson system.  The resulting
colours are for ESHC1 ($U-B=-0.45$, $B-V=0.04$) and for ESHC2
($U-B=-0.62$ and $B-V=0.20$).\\ We apply the seminal Q method (Johnson
\& Morgan 1953) to determine reddening-free colour $Q=(U-B)-0.72
(B-V)$. We derived $Q=-0.47$ for ESHC1 and $Q=-0.76$ for ESHC2.  Using
the relation between $Q$ and spectral type for Be stars by Halbedel
(1993) we derive a spectral type B5 and B2 respectively. This agrees
very well with the spectral types of B4eIII and B2eIV-V derived from
the H and HeI lines above.

A word of caution may be necessary here. The $Q$ versus spectral type
relation that we adopted is derived for Galactic stars. The metal poor
SMC stars, which have less line blanketing and line blocking may have
different colours than Galactic stars of the same $T_{\rm eff}$, and
therefore different temperatures than the Galactic stars of the same
$Q$-value. However, comparing the colours and magnitudes of the Kurucz
model atmospheres (1979) for $Z=0.02$ (Galactic) and $Z=0.002$ (lower
than SMC), we find that the difference is very small.  Given the
uncertainty in the method and the fact that the H and HeI lines
indicate the same type as the $Q$-method, we will adopt these spectral
types and their corresponding values of \Teff\ from Galactic stars.
 
\section{Determination of  the stellar parameters \Teff, $L$, $M$, $R$}
\subsection{ESHC1}

We adopt a spectral type B4 III for ESHC1 as the most reasonable
estimate. From the OGLE photometry the mean magnitude and colours are
$V=15.08\pm 0.01$ and $(B-V)=0.00 \pm 0.03$.  The intrinsic colour of
a B4III star is $(B-V)_0=-0.19$ (Lang 1991), which results in $E(B-V)=
0.19\pm 0.03$.  We make the assumption that the colour difference is
due to obscuration by normal dust with $R_V=3.3$ (Cardelli et al.,
1989). This results in an intrinsic magnitude of $V_0=14.45 \pm 0.11$.
Adopting a distance modulus of $18.94\pm 0.04$ (Laney \& Stobie 1994)
to the SMC, we find $M_V=-4.49\pm 0.15$.  With a bolometric correction
of -1.45 for spectral type B4III (Lang 1991) this yields $M_{\rm
bol}=-5.94\pm 0.15$ or $\log(L/L_\odot)=4.23 \pm 0.06$.  The effective
temperature of a B4III star is log $\Teff =4.20\pm 0.02$ (Lang 1991)
and so $\log(R/R_\odot)= 1.24 \pm 0.03$ and $R=17.4~\pm 1.2~R_\odot$.
Following Lang, we use an empirical Mass Luminosity relation 
$\log (M/M_\odot) = 0.48 -0.105 M_{\rm {bol}}$ and derive a mass of $M\approx 13 M_\odot $.  
Similar results are obtained when we derive the parameters
with the photometry we obtained from ESO telescope.

\subsection{ESHC2}

We adopt a spectral type B2 IV-V for ESHC2 as the most reasonable
estimate.  From the OGLE photometry the mean magnitude and colours are
$V=17.01\pm 0.02$ and $(B-V)=-0.06\pm 0.03$. Adopting the intrinsic
colour $(B-V)_0=-0.24$, an effective temperature of log $\Teff=4.34
\pm 0.02$ and a bolometric correction of $BC=-2.35$ from Lang (1991)
we find the following stellar parameters: $V_0=16.42 \pm 0.12$,
$M_V=-2.52 \pm 0.16$, $M_{\rm bol}=-4.87\pm 0.16$,
$\log (L/L_\odot)=3.84 \pm 0.06$ and $R=6.1 \pm 1 R_\odot$. We get a
mass of $M\approx 10 M_\odot $.  Again, similar results are obtained
when we derive the parameters with the photometry we obtained from the
ESO telescope.  The results are summarized in Table 3.

\begin{table}
 \begin{center}
 \begin{tabular}{lllll}
 \hline
 Name & Type & $\log \Teff$  & $\log(L/L_\odot)$ & $(R/R_\odot)$ \\
 \hline
 ESHC1 & B4 III  & $4.20 \pm 0.02$ & $4.23  \pm 0.06$ & $ 17.4 \pm 1.2$\\
 ESHC2 & B2 IV-V & $4.34 \pm 0.06$ & $3.84 \pm 0.06$ & $\phantom{1}6.1\pm 1$\\
 \hline
 \end{tabular}
 \end{center}
 \caption{Summary of stellar parameters for ESHC1 and ESHC2.}
\end{table}

\section{Discussion}
 
We have presented two stars with the following characteristics:\\ (1)
early B spectral type with strong $H\alpha$ emission, \\ (2) irregular
photometric behaviour with time scales of 20-200 days,\\ (3) a bluer
photometric colour, during brightness minimum,\\ (4) located within a
projected distance of 20\,pc of each other in a weak HII region,\\ (5)
positioned in the HR-diagram to the right of the Main Sequence (see
below).\\ The five points suggest that the stars might be Herbig AeBe or
classical Be stars. If they are PMS stars, then
they are the first discovered Pre-Main Sequence stars in the
SMC (see also Hutchings \& Thompson 1988).

We rule out that the objects are B[e] supergiants, as no forbidden line emission
was detected. Moreover the ESHCs have a lower luminosity class, at
variance with the supergiant nature of B[e] stars in the Magellanic Clouds (Lamers et al. 1998).

\subsection{Comparing the properties of ESHC1 and ESHC2 with classical Be
and HAeBe stars}  

To make a distinction based on visual properties between the two possible
classification is precarious as many properties are common to both
classical Be stars and Herbig Be stars. However we might find some
indications, by having a closer look at the characteristics mentioned
above. The known physical processes (ie. bound-free and free-free
continuum emission, noted bf and ff respectively hereafter) which
occur in the CS environment of classical Be stars allow predictions
with respect to the $H\alpha$ emission and colour excess $E(B-V)$. In
contrast, the activity exhibited by Galactic HAeBe stars remains
fuelling active debate, and therefore no quantitative predictions can
be made.

\noindent - {\bf The $H\alpha$ emission} of Classical Be stars is correlated with the
observed colour excess (Dachs 1988). It is expected that a larger
contribution of the bf-ff excess emission will generate larger
$H\alpha$ emission accompanied by a larger colour excess.  Early type
Galactic classical Be stars have a maximal colour excesses of
$E(B-V)\sim0.10$ (Waters et al. 1987). The derived $E(B-V)$ for
ESHC1 and ESHC2 are 0.19 and 0.18. If we correct for the average
extinction towards the SMC ($E(B-V) =0.09$, Massey et al. 1995), we
find that the colour excesses of ESHC1 and ESHC2 are similar to the
maximum observed for the early type Galactic Be stars (Waters et al.,
1987).  However, comparing the colour excess with the $H\alpha$ 
emission for classical Be stars done by
Dachs (1988) we noticed that the colour excess is much larger than
expected on the basis of the $H \alpha$ emission. This might be
attributed to additional circumstellar extinction.
In general Galactic HAeBe have colour excess which exceeds the
level observed for classical Be stars. 

\noindent - {\bf The irregular brightness variability} of Classical Be stars
has been shown to be largest among spectral types B1-B2, on the basis
of the four years (similar to the EROS2 data) of observations of
Hipparcos (Hubert \& Floquet 1998).  Amplitudes of variability in the
Hipparcos passband were measured to have maximum values of $\Delta
Hp=0.3$, associated with outbursts on time spans of hundreds of days.
This variability is accompanied by small amplitude ($<0.1^{m}$)
variation on time scales of days, often periodic (e.g. Pavlovski et
al. 1997, Hubert \& Floquet 1998). The amplitude of the variability of
the ESHCs is mildly larger the classical Be stars, however the
Hipparcos filter is wider and extends further to the blue part of the
spectrum than the EROS2 $V_{E}$ passband.  Bf-ff emission has its
largest contribution in the red part. Therefore the amplitude in the
EROS2 passband is expected to be mildly larger than the Hipparcos
passband. The superposed short time scale variability of the ESHCs
($\sim 20$\ days), is however not reported for Be stars by Hubert \&
Floquet (1998). This time-scale of variability is however observed
among the HAeBe stars (Herbst \& Shevchenko 1999).\\

\noindent - {\bf The observed colour variability} of the ESHCs can be interpreted
in terms of physical processes, which are thought to occur in the CS
environment of classical Be stars (Dachs 1982, Waters et al. 1987).
However this explanation is not unique. Just like the two SMC HAeBe
candidates, most of the LMC Pre-Main Sequence candidates also show
bluer-when-fainter behaviour. In LBD we discuss the possibility of
explaining the observed photometric and colour variability in terms of
dust obscuration and the existence of a reflection nebula. The
existence of a reflection nebula is one of the original selection
criteria of Herbig (1960). The total flux of HAeBe stars at large
distances (like in the LMC and in the SMC), will be contaminated by
the unresolved, blue reflection nebula (typical sizes of Galactic
reflection nebulae $\sim0.25$\,pc will be less than 1 arcsecond in the
Magellanic Clouds). The nebula will also cause the colour excess to be
small. In LBD it was concluded that one would preferentially observe
Magellanic Cloud HAeBes having a bluer-when-fainter colour behaviour due
to dust obscuration.\\

\noindent - {\bf The close proximity of the two ESHCs} to each other would mildly
support the HAeBe classification. However, we cannot exclude the
chance nearby projection of two irregularly variable classical Be
stars. The H\,II region does indicate the presence of hot stars and
thus a region which might have recently witnessed star formation.

\subsection{Position in the HR-diagram}

We have seen in the previous subsection that the (visual) evidence
does not point towards one particular classification (classical Be or
Herbig Be) of the ESHCs. In the two panels of Fig. 7 we present an HR-diagram. The
purpose is to compare the derived stellar parameters of the ESHCs with
the Classical Be stars and the Herbig Ae/Be stars. The ESHCs are
presented as filled symbols. The Galactic HAeBes are plotted as crosses, and 
the open symbols are their counterpart candidates in the LMC, the ELHCs. 
The full line is the Main-Sequence.  

\begin{figure}
 \centerline{\psfig{figure=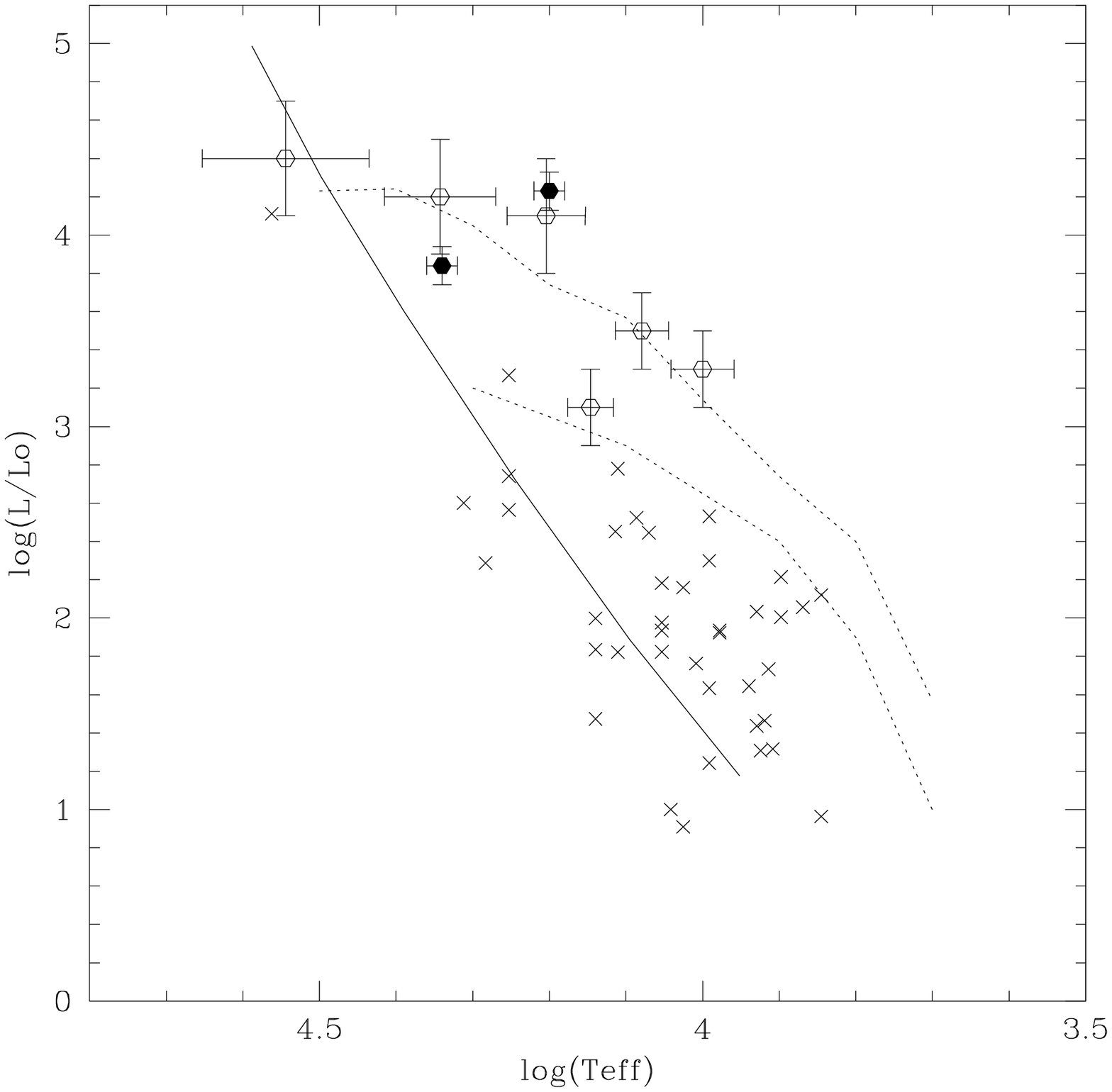,height=9cm,width=9cm}}
 \centerline{\psfig{figure=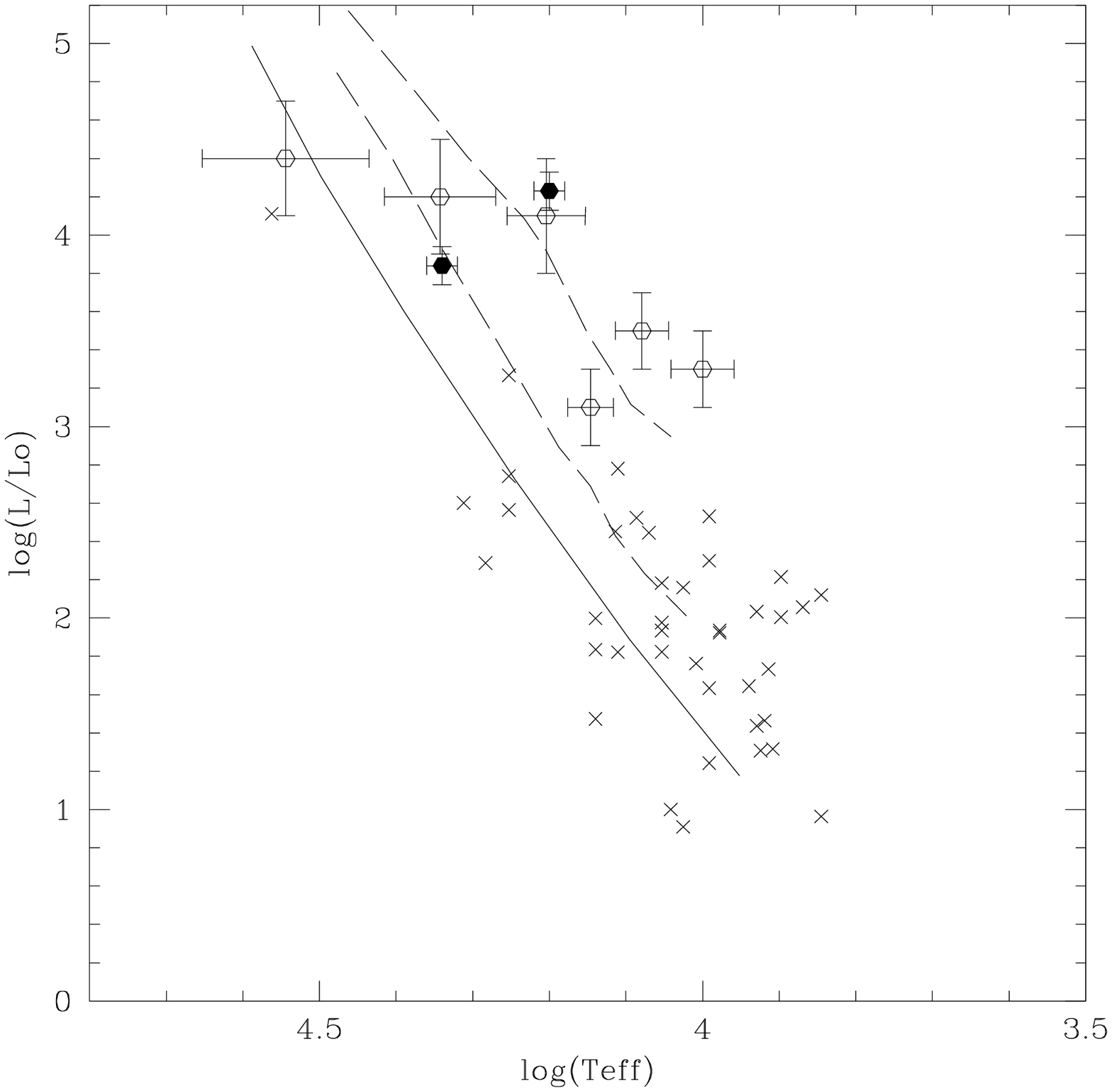,height=9cm,width=9cm}}
 \caption{Both panels are HR diagram for Galactic (crosses, 
Berrilli et al. 1992), LMC (open circles, Lamers et al. 1999) Herbig Ae Be candidates and the two 
targets of this study. The main sequence is indicated as a line. The upper panel shows 
also the two birthlines from Palla \& Stahler (1993) for  mass accretion rates 
 of $10^{-5}M_\odot $ yr$^{-1}$ (long dashed lines) and  
$10^{-4}M_\odot $ yr$^{-1}$ (short dashed lines). The lower panel shows 
(long dashed lines) the upper limit for main sequence Be stars and giant 
Be stars, adopted from Zorec \& Briot (1997).}
\end{figure}

The dotted lines in the upper panel of Fig.\,7 represent the ``birthline'' for pre-Main
Sequence stars for two different proto-stellar mass accretion
rates. The lower dotted line is the canonical Galactic birthline for a
accretion rate of $10^{-5}M_\odot$yr$^{-1}$, and higher dotted line is
the birthline for an accretion rate $10^{-4}M_\odot$ yr$^{-1}$ (Palla \& Stahler
1993). The
crosses are Galactic HAeBe stars adopted from Berrilli et al. (1992).
The observed upper limit for Galactic HAeBes agrees approximately with
an accretion rate of $10^{-5}M_\odot$ yr$^{-1}$.  We see, as noticed
in BL and LBD, that some of the LMC HAeBe candidates (open symbols)
are more luminous by a factor 3 to 10 than the Galactic birthline. In
comparison the luminosity of SMC star ESHC1 complies with the high
luminosity found for the LMC HAeBe candidates.

The luminosity of the birthline depends on the accretion rate. We
found that some of the ELHCs and ESHC1 agree better with the
birthline corresponding to a fast accretion rate of $10^{-4}M_\odot$
$yr^{-1}$. Just based on these two stars, we cannot compare birthlines
in LMC and SMC yet, but it is clear that these two pre-main sequence
candidates in the SMC are above the Galactic birthline, and are very
similar to the LMC pre-main sequence candidates.  {\it If indeed they
are pre-main sequence stars, it would support our previous conclusion
that, at low metallicity, we can observe more massive stars in
pre-main sequence phases than in our Galaxy.}

Palla (2000, private communication) has argued that massive Galactic
PMS may also be above the birthline for an accretion rate of
$10^{-5}M_\odot$yr$^{-1}$. If that is the case, the high location of
ELHCs and ESHCs in the HRD may not be necessarily related to the low
metallicity. On the other hand the massive PMS stars noted by Palla
are closer to the Main Sequence than the ELHCs and ESHCs. So, in the
temperature range $10000<T_{eff}<15000$, the LMC and SMC HAeBe
candidates are definitively above the Galactic HAeBe stars.

On the other hand, we consider the case that they can be classical Be
stars.  The dashed lines in the lower panel of Fig.\,7 indicate the
upper limit for main sequence and giant classical Be stars. These two
upper limits have been derived from Zorec \& Briot (1997). In this
work the average V-Magnitude excess of classical Be stars has been
derived as function of B spectral subtype. We added this excess to the
the absolute visual magnitude (also given in Zorec \& Briot 1997), in
which we included the dispersion $\sigma_{M_{V}}$ in order to obtain
the upper limit. The effective temperature and bolometric correction
were adopted from Schmidt-Kaler (1982).  The figure shows that the
stellar parameters of ESHC2 are compatible with the derived
classical Be upper limits for main sequence stars, and ESHC1 with
the upper limits for giants.

It is not possible to unambiguously distinguish between HAeBe and
classical Be interpretation for the two stars!

\section{Conclusions}
We have identified two blue irregular variables in the SMC with spectral
type B4IIIe (ESHC1) and B2IV-Ve (ESHC2), with strong $H\alpha$ emission.
These stars could be either classical Be stars or pre-main Sequence HAeBe stars.
We find evidence for both interpretations:
\begin{itemize}
\item The $H\alpha$ emission of the ESHCs stars is much larger than that of
the classical Be stars with the same colour excess.
\item The short timescale variability of the ESHC stars of about 20 days is
similar to that of HAeBe stars, but not found in classical Be stars.
\item The bluer-when-fainter colour variability could be explained by variable
free-free and bound-free emission (as in classical Be stars) or by variable
extinction of a star with a blue scattering nebula that is not resolved (as
expected for HAeBe stars at the SMC distance).
\item The presence of a nearby H\,II region and the proximity of clusters of
$H\alpha$-emitters indicates recent star formation, which would favour the HAeBe
interpretation.
\end{itemize}

So we find a mild preference for the interpretation of the ESHC stars as HAeBe
stars. On the other hand, the possibility that the stars are classical Be stars
with abnormal photometric properies cannot be excluded. Infrared observations
and high resolution observations are needed to confirm or deny one of the two
possibilities, because HAeBe stars are known to have an IR excess due to dust
emission and HAeBe stars are often surrounden by a reflection nebula.
 
If  the stars turn out to be HAeBe stars indeed, they are the first pre-main
sequence stars found in the SMC. Their magnitude then indicates that they are
located above the Galactic birthline in the HR-diagram, which suggests a faster
accretion rate than their Galactic counterparts.

\begin{acknowledgements}
This work is based on data obtained at ESO La Silla.
We thank Dr. Zorec for useful discussions and Dr. Grinin for constructive comments on the manuscript. 
S.T. thanks Dr. Belial for methodological approach. This work was partly supported by the Van Gogh program. 
We are grateful to D. Lacroix and the technical staff at the Observatoire de Haute Provence and to A. Baranne for their help in
refurbishing the MARLY telescope and remounting it in La Silla. We are also grateful for the support given to our project by the
technical staff at ESO, La Silla. We thank J. F. Lecointe for assistance with the online computing. 
\end{acknowledgements}


\begin{thebibliography}{}
\bibitem[Alard 2000]{Gnome} {Alard}, C. 2000, in ASP Conf. Ser. 203 (IAU Colloq. 176): The Impact of  Large-Scale Surveys on Pulsating Star Research, 50 
\bibitem[Beaulieu et al. (1996)]{beaulieu} Beaulieu J.P. et al., 1996 Science 272, 995
\bibitem[Berrilli et al. (1992)]{berrilli} Berrilli F., Corciulo G., Ingrosso G., Lorenzetti D., Nisini B., Strafella F., 1992, ApJ 398, 254
\bibitem[Bibo \& Thé (1991)]{bibo} Bibo E.A. \& Th\'{e} P.S. 1991, A\&AS 89, 319
\bibitem[Cardelli et al. (1989)]{} Cardelli J.A., Clayton G.C. \& Mathis J.S. 1989, ApJ 345, 245
\bibitem[{{Dachs}(1982)}]{1982IAUS...98...19D}
{Dachs}, J. 1982, in IAU Symp. 98: Be Stars, Vol.~98, 19
\bibitem[de Wit, W.J.M., Beaulieu J.P. \& Lamers H.J.G.L.M., (2001a)]{} de Wit, W.J.M, Beaulieu J.P. \& Lamers H.J.L.M., (2001a) A\&A submitted
\bibitem[de Wit, W.J.M., Beaulieu J.P. Lamers H.J.G.L.M., Marquette J.B. \& Lesquoy E,  (2001b)]{} de Wit, W.J.M, Beaulieu J.P. Lamers H.J.L.M.,  Marquette J.B. \& Lesquoy E, (2001b) A\&A submitted
\bibitem[Didelon (1982)]{} Didelon, P. 1982 A\&A Suppl., 50, 199
\bibitem[Finkezeller U. \& Mundt R., 1984 ]{finke} Finkezeller U. \& Mundt R., 1984 A\&AS 55, 109 
\bibitem[Halbedel (1973)]{} Halbedel E.M., 1973 PASP 105, 465
\bibitem[Herbig G.H. (1960)]{herbig1} Herbig G.H., 1960, ApJS 4, 337
\bibitem[Herbig G.H. (1994)]{herbig2}  Herbig G.H., 1994,  In : Th\'{e} P.S., Perez M.R., van den Heuvel E.P.J. (eds) The Nature and evolutionnary 
   status of Herbig Ae/Be stars, ASP Conference Series 62., p. 3
\bibitem[{{Herbst} \& {Shevchenko}(1999)}]{1999AJ....118.1043S} {Herbst}, W. \& {Shevchenko}, V.~S. 1999, AJ, 118, 1043
\bibitem[{{Hubert} \& {Floquet}(1998)}]{1998A&A...335..565H} {Hubert}, A.~M. \& {Floquet}, M. 1998, A\&A, 335, 565           
\bibitem[{{Hutchings} \& {Thompson}(1988)}]{1988ApJ...331..294H} {Hutchings}, J.~B. \& {Thompson}, I.~B. 1988, ApJ, 331, 294
\bibitem[Johnson \& Morgan, 1953]{} Johnson H.L. \& Morgan W.W., 1953 ApJ 117, 113
\bibitem[Kurucz (1979)]{} Kurucz R.L., 1979 ApJS 40, 1
\bibitem[Lamers et al. (1999)]{lamers} Lamers H.J.G.L.M., Beaulieu J.P., De Wit W.J., 1999 A\&A 341, 827
\bibitem[Lamers et al. (1998)]{lamers} Lamers H.J.G.L.M., et al., 1998 A\& A 340, L117
\bibitem[Laney \& Stobie (1994)]{} Laney D. \& Stobie R. 1994, MNRAS, 266, 441
\bibitem[ Lang K.R. (1991)]{lang} Lang K.R. 1991, Astrophysical data : planets and stars, Springer-Verlag Berlin p. 114
\bibitem[{{Massey} {et~al.}(1995){Massey}, {Lang}, {Degioia-Eastwood}, \&  {Garmany}}]{1995ApJ...438..188M} {Massey}, P., {Lang}, C.~C., {Degioia-Eastwood}, K., \& {Garmany}, C.~D. 1995,  ApJ, 438, 188                                                                                                                 
\bibitem[Meysonnier N. \& Azzopardi M., (1993)]{Azzo} Meysonnier N. \& Azzopardi M. 1993,  A\&A Suppl. 102, 451
\bibitem[Palla F. \& Stahler S.W. (1993)]{palla} Palla F. \& Stahler S.W. 1993, ApJ 418, 414
\bibitem[Palanque-Delabrouille et al. (1998)]{nath}  Palanque-Delabrouille N. et al. (EROS Collaboration) 1998, A\&A 332, 1
\bibitem[{{Schmidt-Kaler}(1982)}]{schmidt-kaler82} {Schmidt-Kaler}, T. 1982, in Landolt Bornstein Vol. 2, Subvol B, Springer  Verlag, Berlin                                            
\bibitem[Thé P.S. (1994)]{the} Th\' {e}.S., 1994, In : Th\'{e} P.S., Perez M.R., van den Heuvel E.P.J. (eds) The Nature and evolutionnary  status of Herbig Ae/Be stars, ASP Conference Series 62., p. 23 
\bibitem[Waters L., Walkens C.,  (1998)]{} Waters L.B.F.M., Waelkens C., 1998 ARAA 36, 233
\bibitem[{{Waters} {et~al.}(1987){Waters}, {Cote}, \&  {Lamers}}]{1987A&A...185..206W}{Waters}, L. B. F.~M., {Cote}, J., {Lamers}, H. J. G. L.~M. 1987, A\&A 185,  206      
\bibitem[{{Zorec} \&  {Briot}}]{1987A&A...185..206W} {Zorec}, J. \&  {Briot}, D. 1997, A\&A 318, 443          

\end{thebibliography}
\end{document}